\documentclass[reprint,twocolumn]{revtex4-1}
\usepackage[centertags]{amsmath}
\usepackage{graphicx}
\usepackage{amsfonts}
\usepackage{color}
\newcounter{defcounter}
\begin{document} 
\title{Modulation Instability and Phase-Shifted Fermi-Pasta-Ulam Recurrence}
\author{O. Kimmoun$^{1\ast}$, H.C. Hsu$^{2}$, H. Branger$^{1}$, M.S. Li$^{2}$,
Y.Y. Chen$^{2}$, C. Kharif$^{1}$, M. Onorato$^{3}$, E. J. R. Kelleher$^{4}$,
 B. Kibler$^{5}$, N. Akhmediev$^{6}$, and A. Chabchoub$^{7}$}
\email{Corresponding author: olivier.kimmoun@centrale-marseille.fr}
\affiliation{$^1$ Aix-Marseille University, CNRS, Centrale Marseille, IRPHE, Marseille, France}
\affiliation{$^2$ Tainan Hydraulics Laboratory, National Cheng Kung University, Taiwan}
\affiliation{$^{3}$Dipartimento di Fisica Generale, Universita degli Studi di Torino, Torino, Italy}
\affiliation{$^{4}$Femtosecond Optics Group, Department of Physics, Imperial College London, London, U.K.}
\affiliation{$^{5}$Laboratoire Interdisciplinaire Carnot de Bourgogne, UMR 6303 CNRS – UBFC, Dijon, France} 
\affiliation{$^{6}$Optical Sciences Group, Research School of Physics and Engineering, Institute of Advanced Studies, The Australian National University, Canberra ACT 020, Australia}
\affiliation{$^{7}$Department of Ocean Technology Policy and Environment, Graduate School of Frontier Sciences, The University of Tokyo, Kashiwa, Chiba 277-8563, Japan}
\begin{abstract}
Instabilities are common phenomena frequently observed in nature, sometimes leading to unexpected catastrophes and disasters in seemingly normal conditions. The simplest form of instability in a distributed system is its response to a harmonic modulation. Such instability has special names in various branches of physics and is generally known as modulation instability (MI). The MI is tightly related to Fermi-Pasta-Ulam (FPU) recurrence since breather solutions of the nonlinear Schr\"odinger equation (NLSE) are known to accurately describe growth and decay of modulationally unstable waves in conservative systems.  Here, we report theoretical, numerical and experimental evidence of the effect of dissipation on FPU cycles in a super wave tank, namely their shift in a determined order. In showing that ideal NLSE breather solutions can describe such dissipative nonlinear dynamics, our results may impact the interpretation of a wide range of new physics scenarios.
\end{abstract}
\maketitle 

\section{INTRODUCTION}
The discovery of the Fermi-Pasta-Ulam (FPU) recurrence was a significant step in nonlinear dynamics. It describes the natural return cycle of a dynamical system to its initial conditions after undergoing complex motion dynamics {\color{blue}\cite{FPU,OnoratoPNAS}}. Meanwhile, the FPU recurrence has been studied and observed in several nonlinear media. For instance, in hydrodynamics within the framework of the Korteweg De Vries {\color{blue}\cite{Zabusky}} equations as well as in a more broad range in physics within the context of the nonlinear Schr\"odinger equation (NLSE) {\color{blue}\cite{Newell,Zakharov}}, particularly, in the description of modulationally unstable periodic packets returning to the initial state of small perturbation of the background after significant envelope compression {\color{blue}\cite{YuenLake1982,Waseda}}. 
In fiber optics, experimental observations of the FPU recurrence have been also restricted to one or two cycles {\color{blue}\cite{VanSimaeys,Kibler2}}. 

More recently, the effects of different perturbations to the standard NLSE such as third-order dispersion or varying dispersion, or even the impact of initial excitation of modulation instability, on the FPU phenomenon have been reported as well {\color{blue}\cite{Mussot,Bendahmane2,Erkintalo2}}.

Here, we show that the impact of weak dissipation engenders shifted FPU recurrence in the localizations of periodic breathers. The results are in very good agreement with laboratory experiments, which have been performed in a super water wave tank. Due to the interdisciplinary character of the approach, this study emphasizes a wide range of applications in other nonlinear dispersive media, as will be discussed. 

\section{THEORETICAL BACKGROUND}
\begin{figure}[ht]
\centering
\includegraphics[width=0.5\columnwidth]{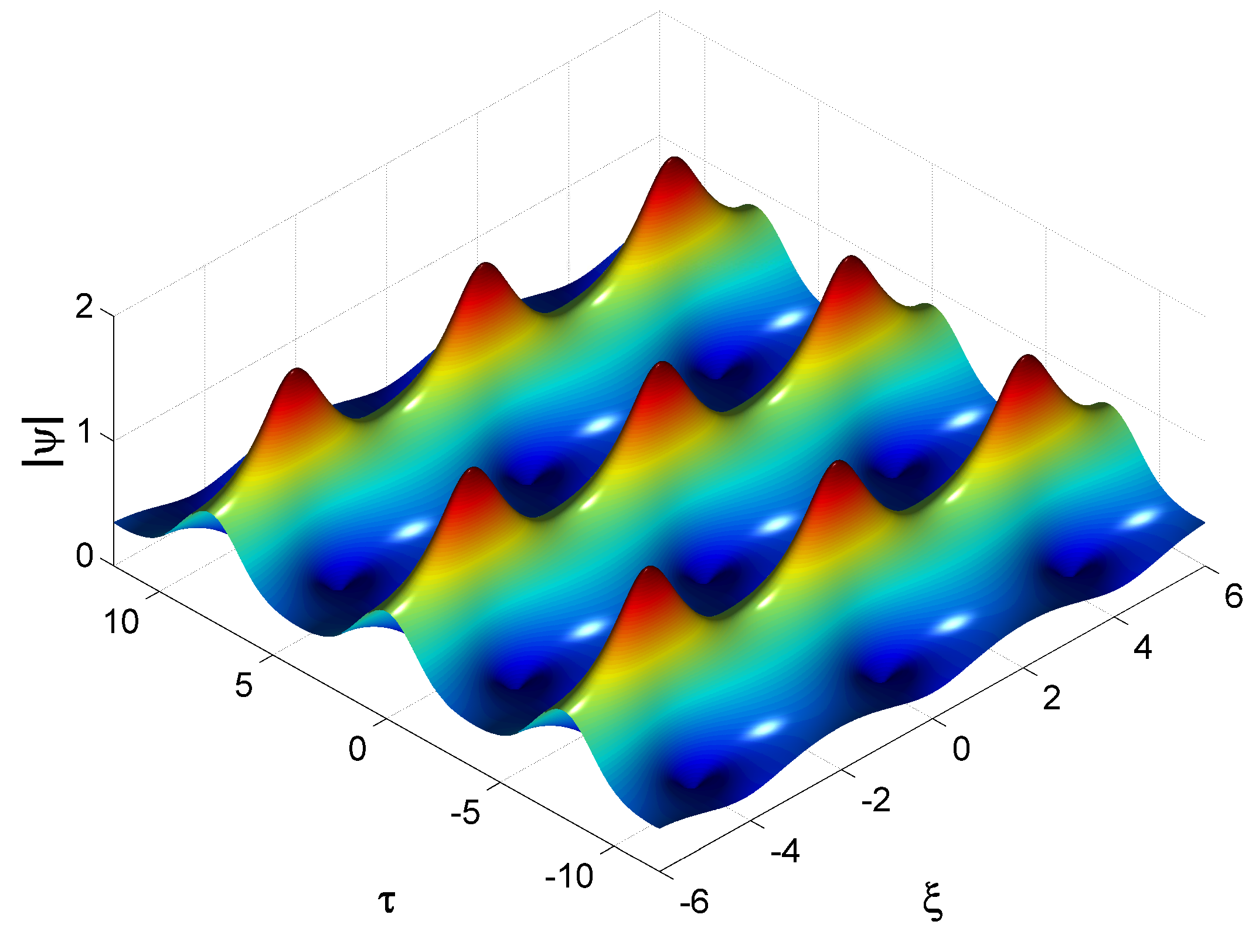}
\caption{Periodic solution (\ref{45}) for $\kappa=0.7$.}
\label{perd1}
\end{figure}

\begin{figure}[ht]
\centering
\includegraphics[width=0.5\columnwidth]{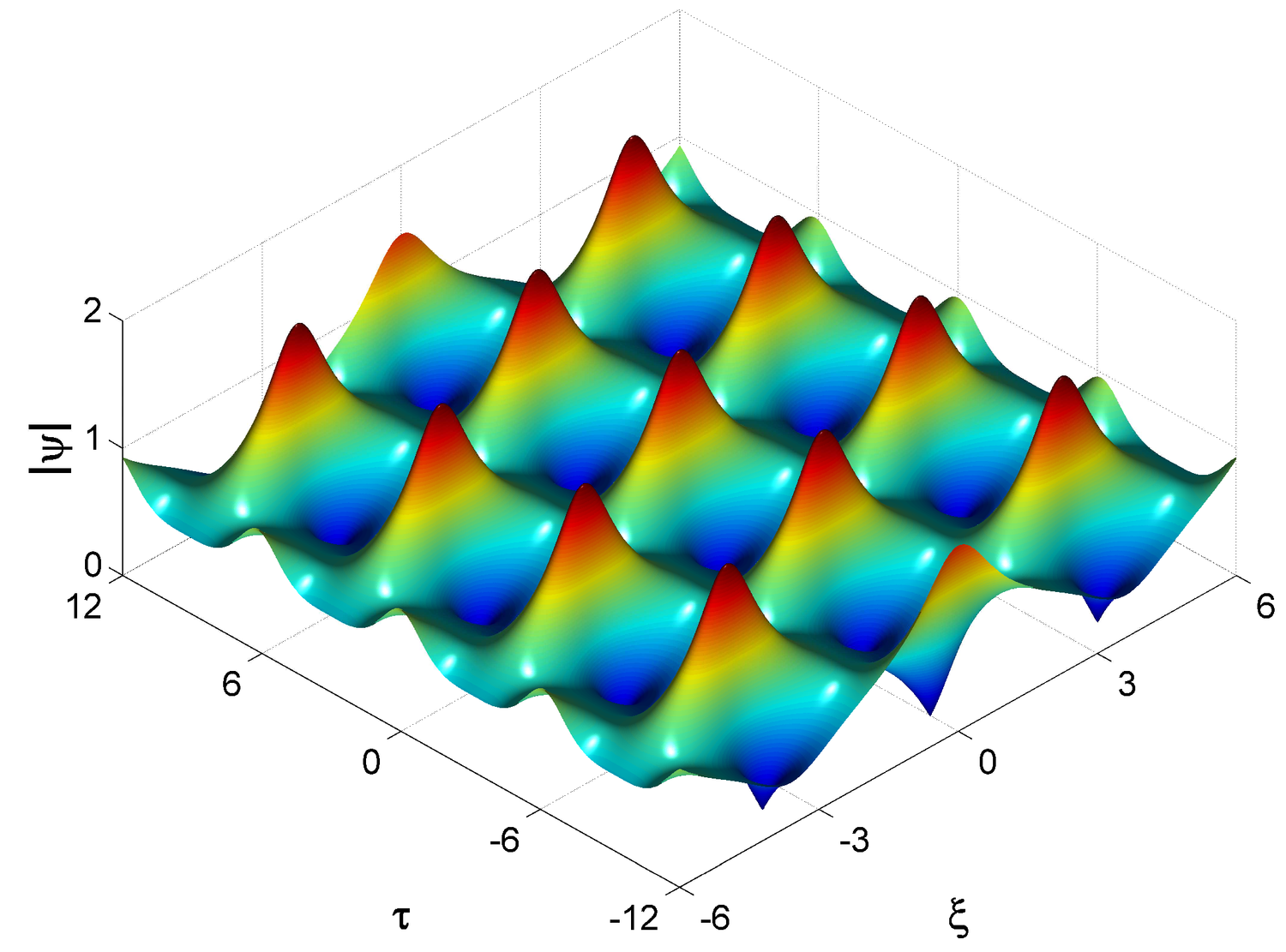}
\caption{periodic solution (\ref{per}) for $k=0.8$.}
\label{perd2}
\end{figure}

The NLSE is one of the most significant equations in physics. It describes waves on the surface of the ocean, pulses in optical fibres, special states of Bose-Einstein condensates, plasma oscillations and many other phenomena. The NLSE can be written in dimensionless form as:
\begin{equation}
\operatorname{i}\psi_{\xi}+\psi_{\tau\tau}+2|\psi|^2\psi=0
\end{equation}
Here, $\xi$ describes the spatial co-ordinate, moving with the group velocity, while $\tau$ denotes the scaled time and $\psi$ the scaled envelope amplitude. 
Among elementary solutions of the NLSE are plane waves, solitons, breathers and rational solutions {\color{blue}\cite{AkhmedievBook}}. Breathers can be considered as heteroclinic orbits connecting two saddle points in an infinite-dimensional phase space. The latter are plane wave solutions that are modulationally unstable. Saddle points are usually surrounded by nearby trajectories that are connected to similar trajectories around the second saddle point. As a result, homoclinic orbits are surrounded by the periodic trajectories and corresponding solutions of the NLSE {\color{blue}\cite{Akhmediev1}}. One of such periodic solution of the NLSE has the form
\begin{equation}\label{45}
\psi(\tau,\xi)=\frac{\kappa}{\sqrt{2}}\,\frac{A(\tau)~\mbox{cn}(\xi,\kappa)+\operatorname{i}\sqrt{1+
\kappa}~\mbox{sn}(\xi,\kappa)}{\sqrt{1+\kappa}-A(\tau)~\mbox{dn}(\xi,\kappa)}\exp\left(\operatorname{i}\xi\right)
\end{equation} 
where
\begin{equation}
A(\tau)=\frac{\mbox{cn} \left[\sqrt{(1+\kappa)/2}\tau,~\sqrt{(1-\kappa)/(1+
\kappa)} \right] }{\mbox{dn} \left[ \sqrt{(1+\kappa)/2}\tau,~\sqrt{(1-\kappa)/(1+\kappa)} \right] } 
\end{equation}
and $\kappa$ is a free parameter of this family of solutions.
This solution is periodic in both $\tau$ and $\xi$ with periods defined by $\kappa$.  It is
illustrated, for the case $\kappa=0.7$, in Fig. \ref{perd1}. 
In the limit $\kappa \rightarrow 1$, it  reduces to the AB solution {\color{blue}\cite{Akhmediev1}} 
\begin{equation} \label{mxr}
\psi(\tau,\xi)=\frac{\cos\left(\tau\right)+\operatorname{i}\sqrt{2}~\mbox{sinh}(\xi)}{\sqrt{2}~\cos\left(\tau\right)-2~\mbox{cosh}(\xi)}\exp\left(\operatorname{i}\xi\right)
\end{equation} 
which has infinite the period in $\xi$ and which is the above mentioned homoclinic orbit.
On the other hand, in the limit $\kappa \rightarrow 0$, period in $\tau$ becomes infinite and it  converges to the well-known basic bright
soliton solution:  
\begin{equation}
\psi_S(\tau,\xi)=\operatorname{sech}\left(\tau\right)\exp\left(\operatorname{i}\xi\right)
\label{sss}
\end{equation}
The periodic solution that is located on the other side of the homoclinic orbit (or separatrix)
 has the form \begin{equation}
\label{per}
\psi(\tau,\xi)=\frac{\Lambda~\mbox{dn}\big{(}\frac{\xi}{k},k\big{)}
+\operatorname{i}\,k~\mbox{sn}\big{(}\frac{\xi}{k},k\big{)}}{k\sqrt{2}\Big{[}1-\Lambda~\mbox{cn}\big{(}\frac{\xi}{k},k\big{)} \Big{]}}\exp\left(\operatorname{i}\xi\right) \end{equation}
with 
$$\footnotesize \Lambda=\sqrt{\frac{k}{1+k}}~\mbox{cn}\Big{(}\frac{\tau}{\sqrt{k}},\sqrt{\frac{1-k}{2}}\,\Big{)}$$
This periodic solution is shown in Fig. \ref{perd2} for $k=0.85$.

It can be alternatively derived from the solution (\ref{45}) taking $\kappa>1$ and using the transformations for
elliptic Jacobi functions for the case $\kappa=1/k$. The latter can be found in {\color{blue}\cite{Abramowitz}}. As $k
\rightarrow 1$, the solution (\ref{per}) has formula (\ref{mxr}) as its limit. As solutions
(\ref{per}) and (\ref{45}) are located on different sides of the separatrix (\ref{mxr}).
they are qualitatively different. The solution (\ref{per}) keeps maxima of the periodic function at the
same position while the solution (\ref{45}) has maxima alternating. The latter can be considered as
phase-shifting solutions. 
This difference can be seen clearly by comparing Fig. \ref{perd2} with Fig. \ref{perd1}.

\begin{figure}[ht]
\centering
\includegraphics[width=\columnwidth]{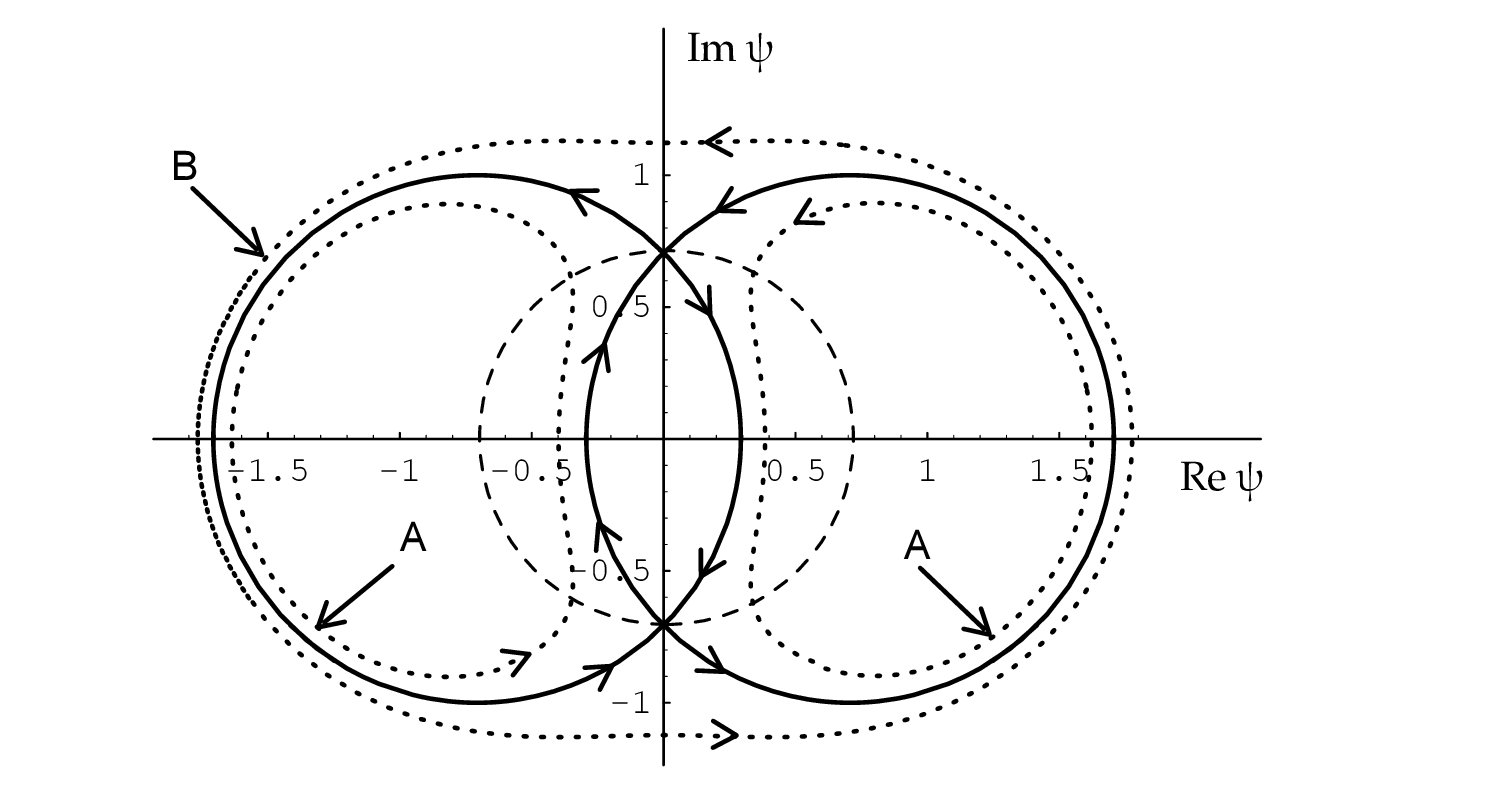}
\caption{Trajectories (bold circles) in the complex plane describing
the modulation instability with the highest initial growth rate. The circle around the origin
(dashed curve) is the manifold of initial conditions. Here $\kappa=1$, i.e.
$a_1=1/4$. The dotted lines schematically show two qualitatively different
types of periodic solution close to the separatrix. The curves labelled A correspond to 
solution (\ref{45}), while  curve B corresponds to (\ref{per}).}
\label{Evol}
\end{figure}

The geometric interpretation of two types of periodic solutions is presented in Fig. \ref{Evol}.
The circle with unit radius shown by the dashed line can be considered as a plane wave with unit amplitude
and variable phase. Each point of the circle is an unstable saddle point. 
The starting trajectories of the saddle point describe Benjamin-Feir or modulation instability (MI). 
Let us choose one of them
located, say, at the upper point of the circle. Continuation of the trajectory that starts at the saddle point
ends up at another saddle point located at the bottom of the circle. 
This particular trajectory describes the Akhmediev breather
(\ref{mxr}). This trajectory is a heteroclinic orbit separating two trajectories denoted as A and B corresponding to 
periodic solutions (\ref{per}) and (\ref{45}). Trajectory A rotates on one side of the
complex plane and does not shift the phase while trajectory B rotates around the origin thus gaining the phase
difference $2\pi$ on each period of oscillations.
\section{FERMI-PASTA-ULAM RECURRENCE AND BREATHERS IN THE PRESENCE OF DISSIPATION}
\begin{figure}[ht]
\centering
\includegraphics[width=\columnwidth]{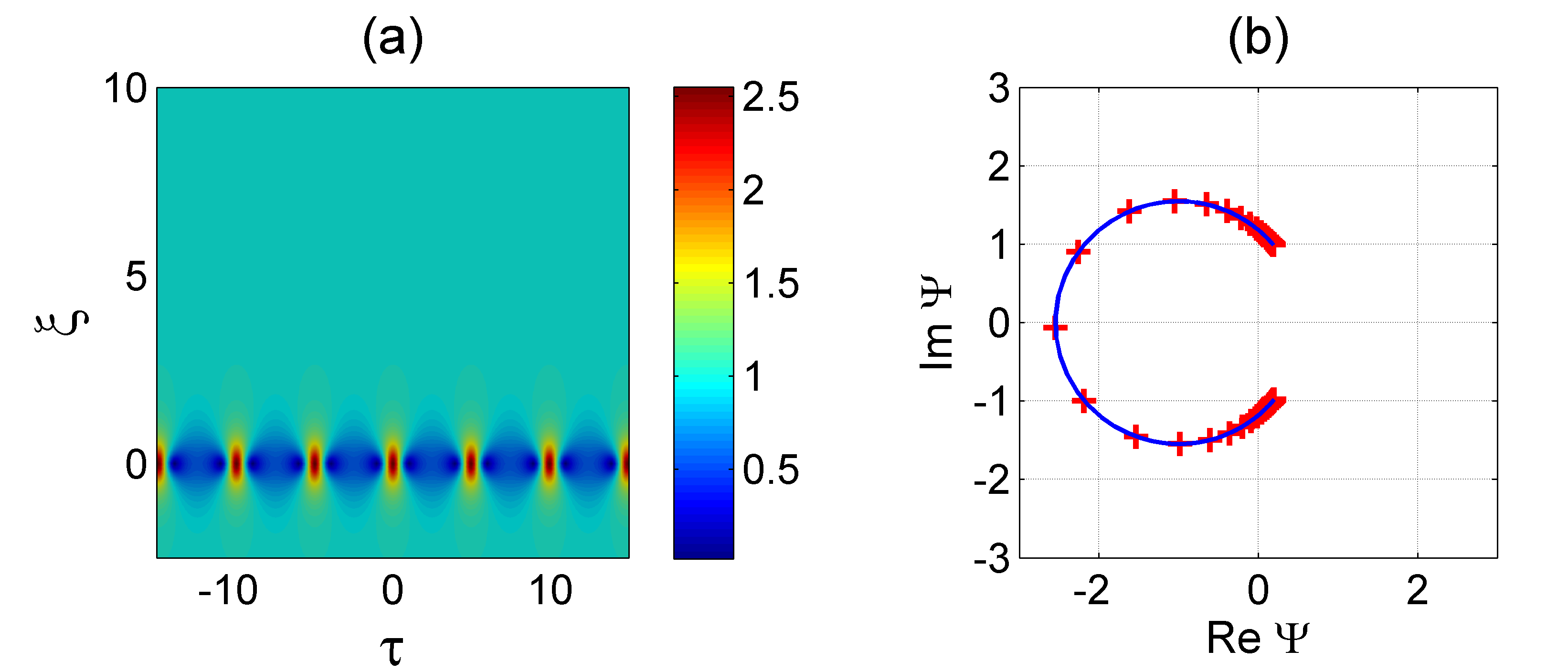}
\caption{(a) Numerical simulation of AB dynamics from NLSE, propagating in space for $\mathfrak{a}=0.3$, and starting from the saddle point given by theory at $\xi = -2.5$. (b) Corresponding trajectory in the complex plane. The red crosses correspond to numerical simulations, while the blue curve is given by the exact analytical expression (\ref{AB}).}
\label{fig4}
\end{figure}

\begin{figure}[ht!]
\centering
\begin{tabular}{c}
\includegraphics[width=\columnwidth]{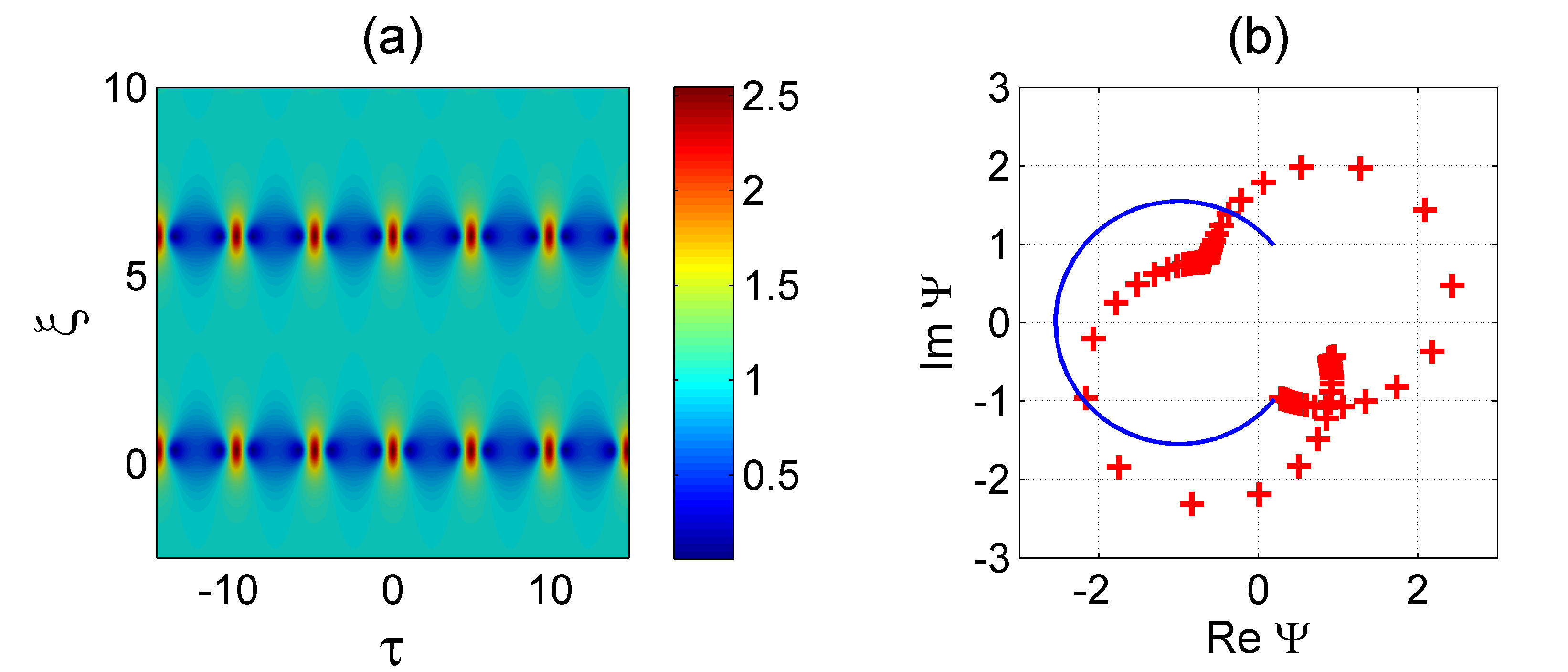}\\
\includegraphics[width=\columnwidth]{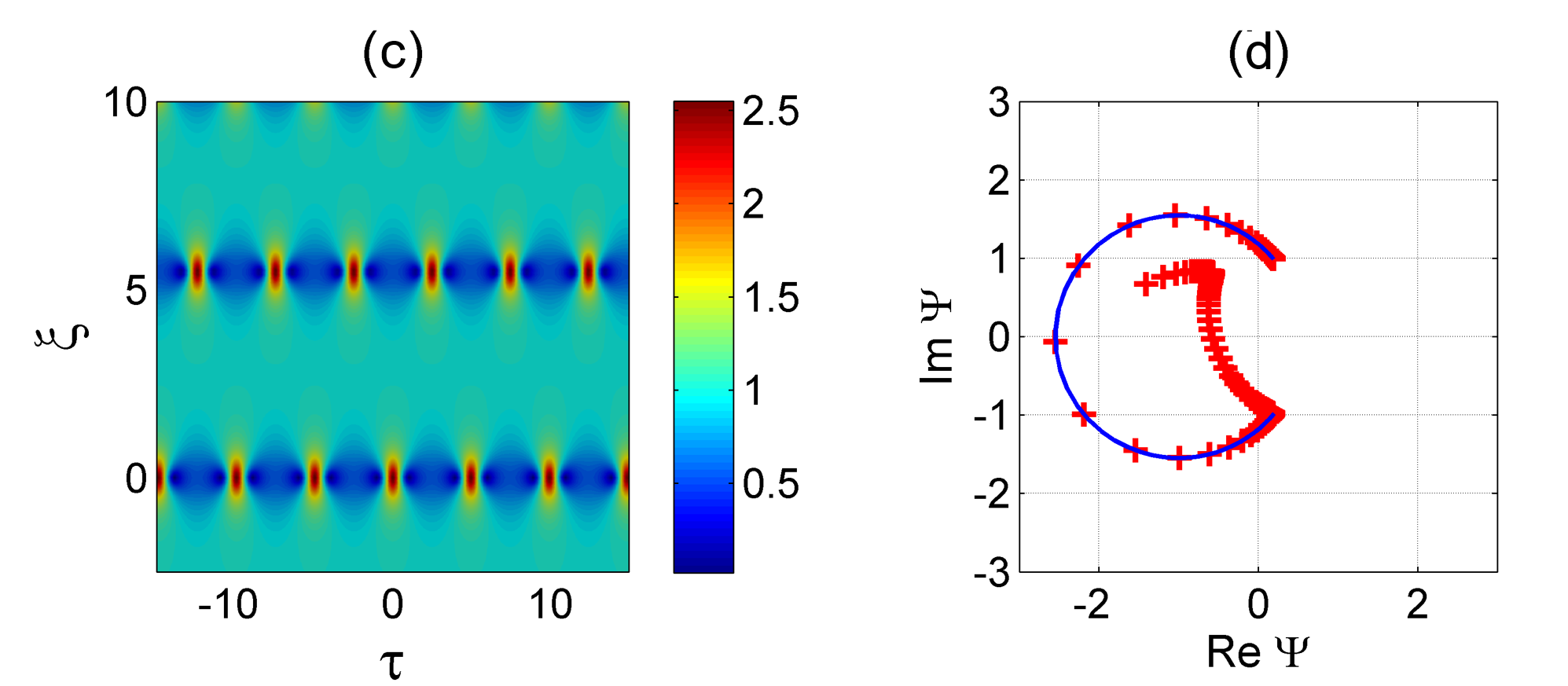}
\end{tabular}
\caption{Numerical simulation of AB dynamics from NLSE, propagating in space for $\mathfrak{a}=0.3$, (a) starting from an approximate cosine modulation of the background wave that fits the theoretical AB profile at $\xi = -2.5$. (c) starting from the theoretical AB profile at $\xi = -2.5$ and using a dissipation rate $\mathfrak{D} = 4.17\,\mbox{x}\,10^{-5}$. (b,d) Corresponding trajectory in the complex plane for the case (a,c) respectively. The red crosses correspond to numerical simulations, while the blue curve is given by the exact analytical expression (\ref{AB}).}
\label{fig5}
\end{figure}

These two types of solutions can be observed in experiments that start with modulation instability. In the presence of even small perturbations, the separatrix may be converted into a nearby periodic orbit of either type A or type B depending on the sign of the perturbation. In the experiments that are described in the present work, the dissipation played the role of the perturbation.
An interesting point here is the fact that despite the dissipation always has the same sign, the perturbation it causes can be either positive or negative. Consequently, the trajectory could be converted either to type A or type B during the evolution. These two scenarios can be easily detected experimentally observing the phase shift.
\\
We confirm the occurrence of this interesting phenomenon in numerical simulations and validate the results through hydrodynamic laboratory experiments conducted in a large wave facility permitting the measurement of several stages of envelope compressions. 
Namely, we reveal that deviations from exact AB envelope dynamics caused by non-ideal excitation or propagation losses imply the emergence of successive spatial recurrences as shown in Figs. \ref{perd1} and \ref{perd2}. In particular, the specific phase-shift of the envelope modulation cycles caused by dissipation is observed experimentally.
\\
The solutions given above are related to the modulation instability with the highest growth rate.
In this particular case, the phase shift of one growth - decay cycle is equal to $\pi$. Any other frequency within the instability band produces its own phase shift which varies from zero to $2\pi$. The whole family of Akhmediev breathers (ABs) {\color{blue}\cite{Akhmediev1}} is given by
\begin{equation}
\begin{array}{ll}
\psi\left(\tau,\xi\right) & = \left[\dfrac{\sqrt{2\mathfrak{a}}\cos\left(\Omega\tau\right)+\left(1-4\mathfrak{a}\right)\cosh\left(R\xi\right)}
{\sqrt{2\mathfrak{a}}\cos\left(\Omega\tau\right)-\cosh\left(R\xi\right)}\right.\\
& +\left. \dfrac{\textnormal{i}R\sinh\left(R\xi\right)}{\sqrt{2\mathfrak{a}}\cos\left(\Omega\tau\right)-\cosh\left(R\xi\right)}\right]
\exp\left(\textnormal{i}\xi\right)
\label{AB}
\end{array}
\end{equation}
where $\Omega=2\sqrt{1-2\mathfrak{a}}$ and $R=\sqrt{2\mathfrak{a}(1-2\mathfrak{a}^2)}$. Here $0<\mathfrak{a}<0.5$
while $\Omega$ and $R$ determine the modulation frequency and corresponding growth or decay rate near the saddle point, respectively. Note that Eq. (\ref{mxr}) is a particular case of the AB solutions (\ref{AB}). When $\mathfrak{a}=0.25$ the solution describes the modulation instability in the event of maximal growth rate. When $\mathfrak{a}>0.5$, the same solution describes the family of space periodic Kuznetsov-Ma breathers {\color{blue}\cite{Kuznetsov,Ma}}.
In the limiting case of $\mathfrak{a} \rightarrow 0.5$, the growth rate becomes algebraic and the solution reduces to a rational doubly-localized solution, known as the Peregrine solution {\color{blue}\cite{Peregrine}}. The latter breather solutions attracted scientific interest recently {\color{blue}\cite{OnoratoReview,DudleyReview}} and had been observed in several nonlinear systems {\color{blue}\cite{DudleyKibler,Kibler,Chabchoub1,Bailung,Kibler2,Chabchoub2}}. Solutions (\ref{AB}) are of significant importance in understanding the modulation instability as well as for a wide range of applications, since they exactly and physically describe the complete growth - decay cycle of modulation instability, quantifying in detail the side-band cascade dynamics {\color{blue}\cite{Mussot,Hammani}}.
\\
Fig. \ref{fig4}(a) shows an example of exact AB evolution for the breather parameter $\mathfrak{a}=0.3$, while Fig. \ref{fig4}(b) shows the trajectory on the complex plane that corresponds to this evolution. 
The total phase shift provided by this trajectory is higher than $\pi$ which means that the MI frequency is detuned from the highest growth rate regime.
Small perturbation of the initial conditions engenders the recurrent breathing of AB envelopes. The trajectory misses exact saddle point and continues along the hyperbolic orbit. 
This results in the next recurrent dynamics which then continues into periodic motion{\color{blue}\cite{Mussot}}. An example of two successive recurrences is depicted in Fig. \ref{fig5}(a) when approximate initial conditions are used in the numerical simulation. Here the theoretical AB profile at $\xi=-2.5$ was fitted by a simple cosine modulation of the background wave with same frequency and amplitude.
In physical systems deviations from exact conditions may be caused by several sources, such as higher-order dispersion and inelastic Raman scattering in optical fibers {\color{blue}\cite{Agrawal}} or the mean flow of Stokes waves in hydrodynamics {\color{blue}\cite{Dysthe}}. Another example is the effect of weak dissipation. The latter shifts the wave profile from the exact shape leading the trajectory to miss the saddle point. The model describing the weak attenuation of the wave envelope is the NLSE with dissipation {\color{blue}\cite{Hasegawa}}:
\begin{eqnarray}
\textnormal{i}\psi_\xi+\psi_{\tau\tau}+2\left|\psi\right|^2\psi=-\textnormal{i}\mathfrak{D}\psi
\end{eqnarray} 
where $\mathfrak{D}$ is the normalized attenuation rate. The effect of linear attenuation on the AB dynamics in particular, can now be studied numerically. As we will see next, the effect of the dissipation will engender a $\pi/2$ phase-shift in the recurrent breather compression. 
Fig. \ref{fig5}(c) shows the corresponding impact on the AB evolution using exact initial conditions, determined by $\mathfrak{a}=0.3$, while the dissipation rate is set to $\mathfrak{D} = 4.17\,\mbox{x}\,10^{-5}$.
\\
This is a remarkable phenomenon, which we will refer to as phase-shifted FPU recurrence.
\section{EXPERIMENTAL SETUP} 
\begin{figure}[ht]
\centering
\includegraphics[width=\columnwidth]{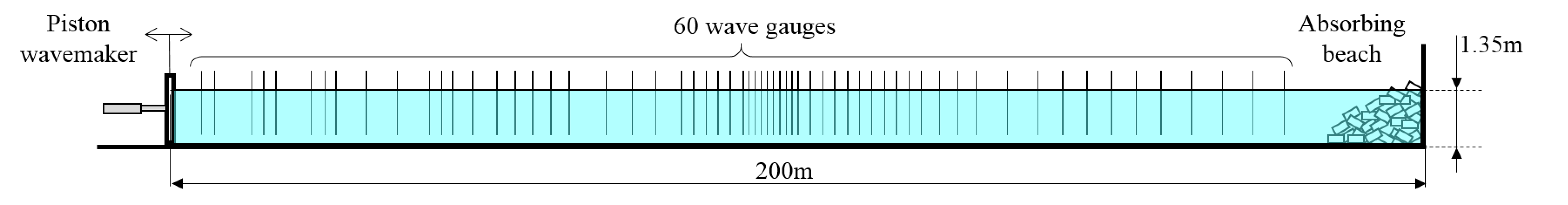}
\caption{Schematic description of wave facility.} 
\label{fig7}
\end{figure}

Experiments have been performed in a super tank, installed at the Tainan Hydraulics Laboratory (THL) of National Cheng Kung University in Taiwan. The facility is 200 m long, 2 m wide and 2 m high. The water depth was set to 1.35 m. The tank is equipped with a piston wave-maker, which generates the waves at one end of the flume, while an absorbing beach is installed at the other end. In order to measure wave elevation, 60 capacitance-type wave gauges, with a sample rate of 100 Hz, have been deployed along the tank and calibrated accordingly, before conducting the experiments. The first gauge was fixed at 2.1 m from the wave-maker, while the last at 176.1 m. A schematic illustration of the facility is shown in Fig. \ref{fig7}. 
\section{EXPERIMENTAL OBSERVATIONS}
\begin{figure}[ht!]
\centering
\begin{tabular}{cc}
\includegraphics[width=0.5\columnwidth]{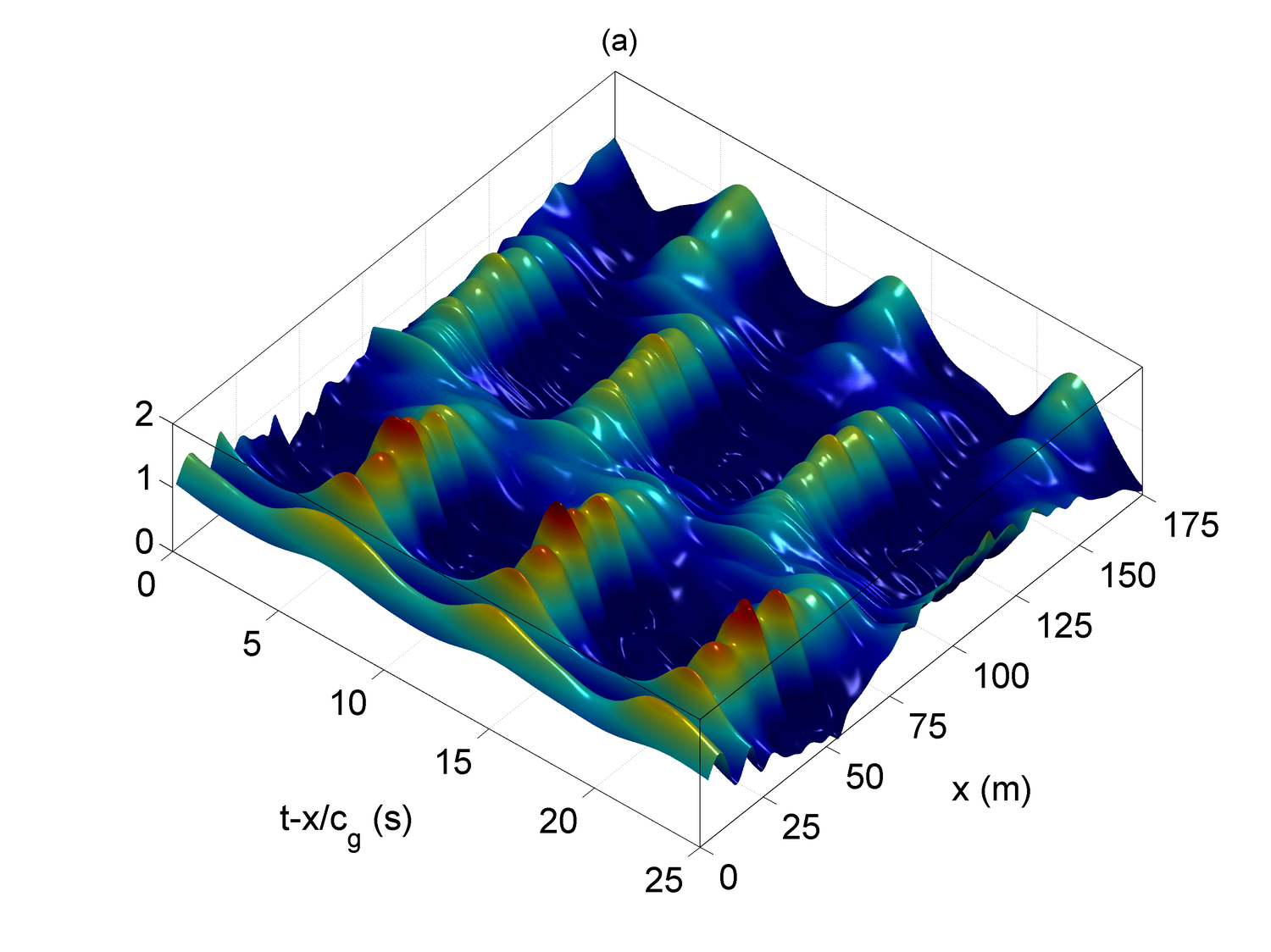}&\includegraphics[width=.5\columnwidth]{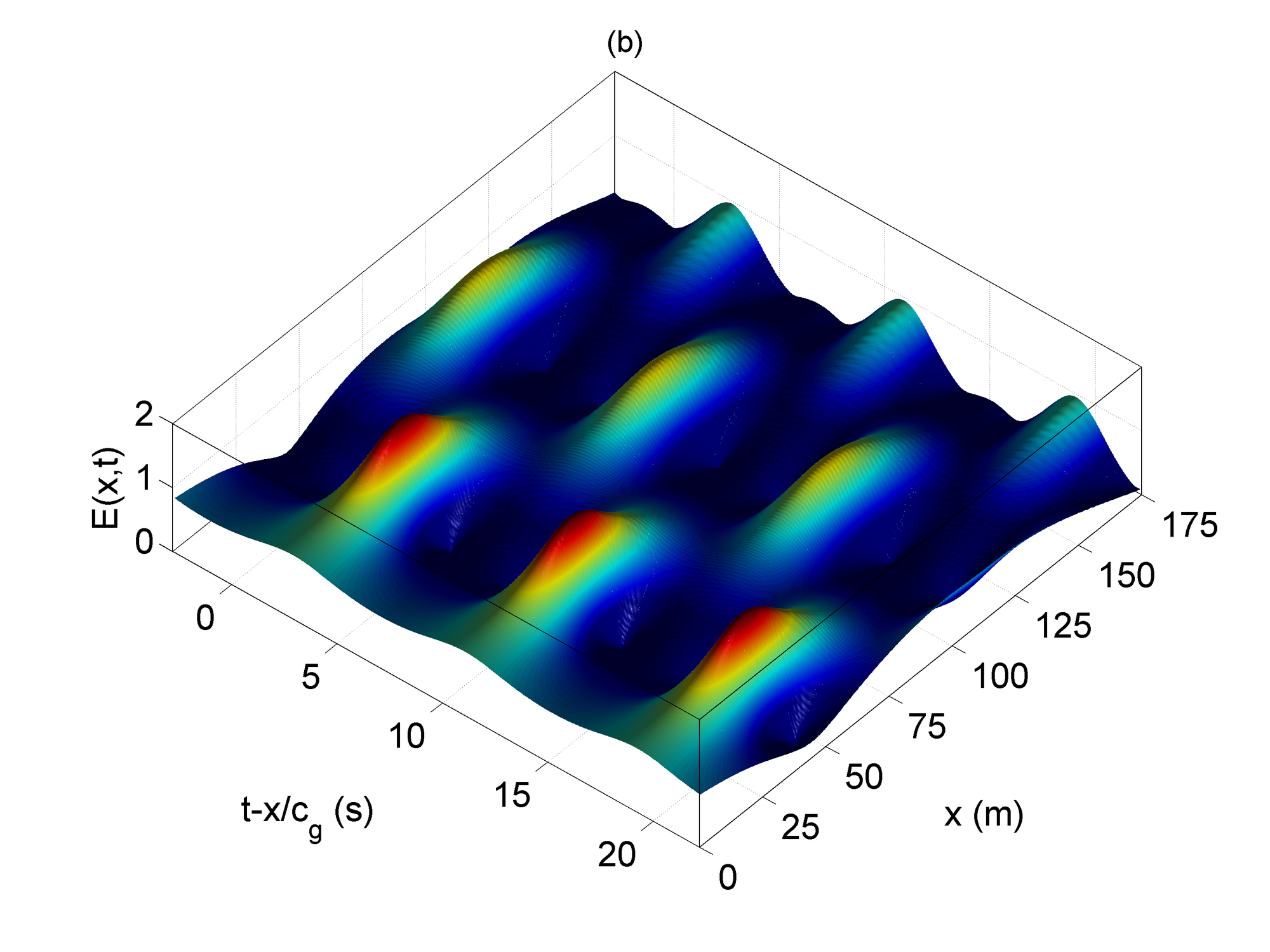}\\
\includegraphics[width=0.5\columnwidth]{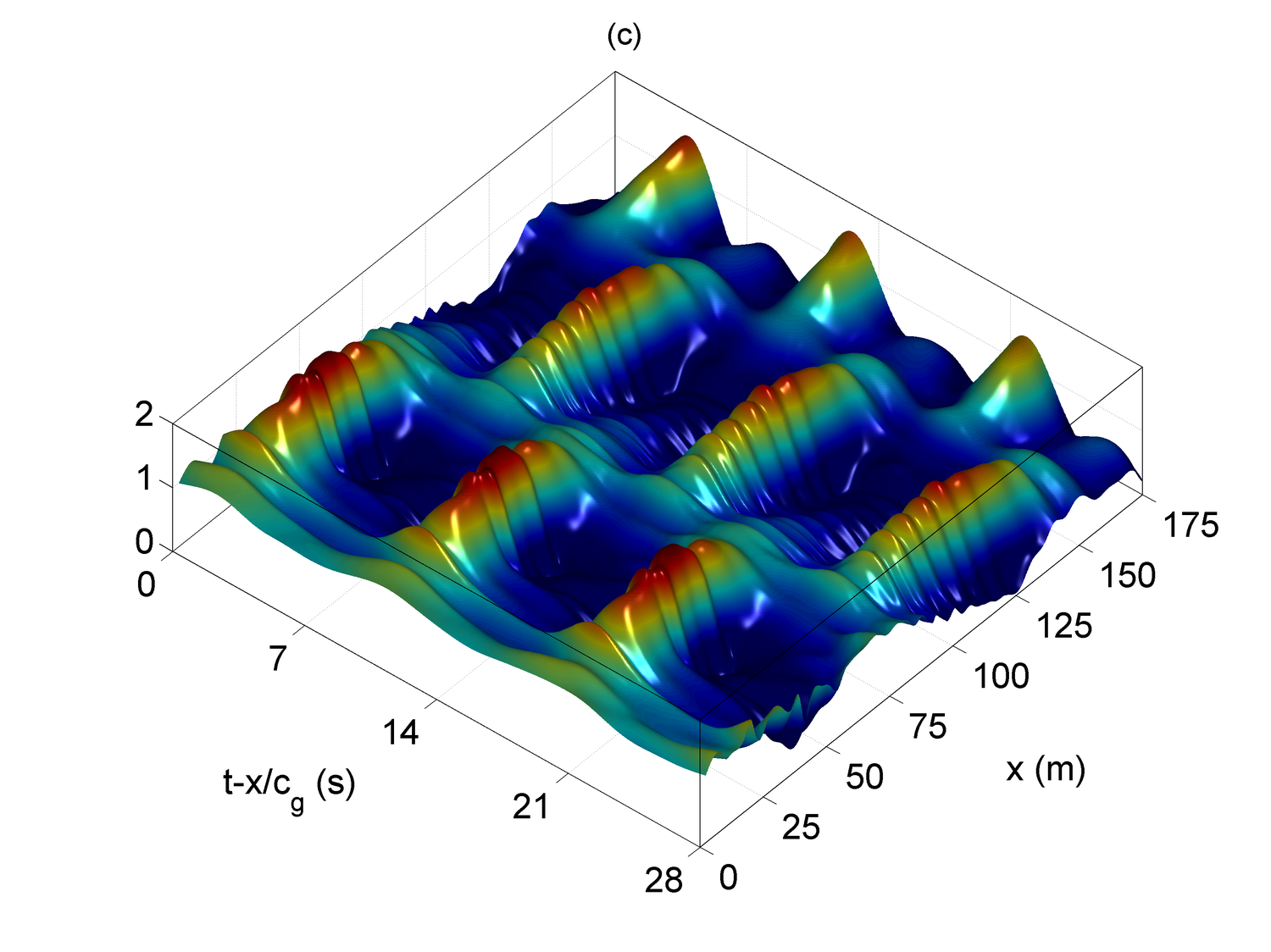}&\includegraphics[width=.5\columnwidth]{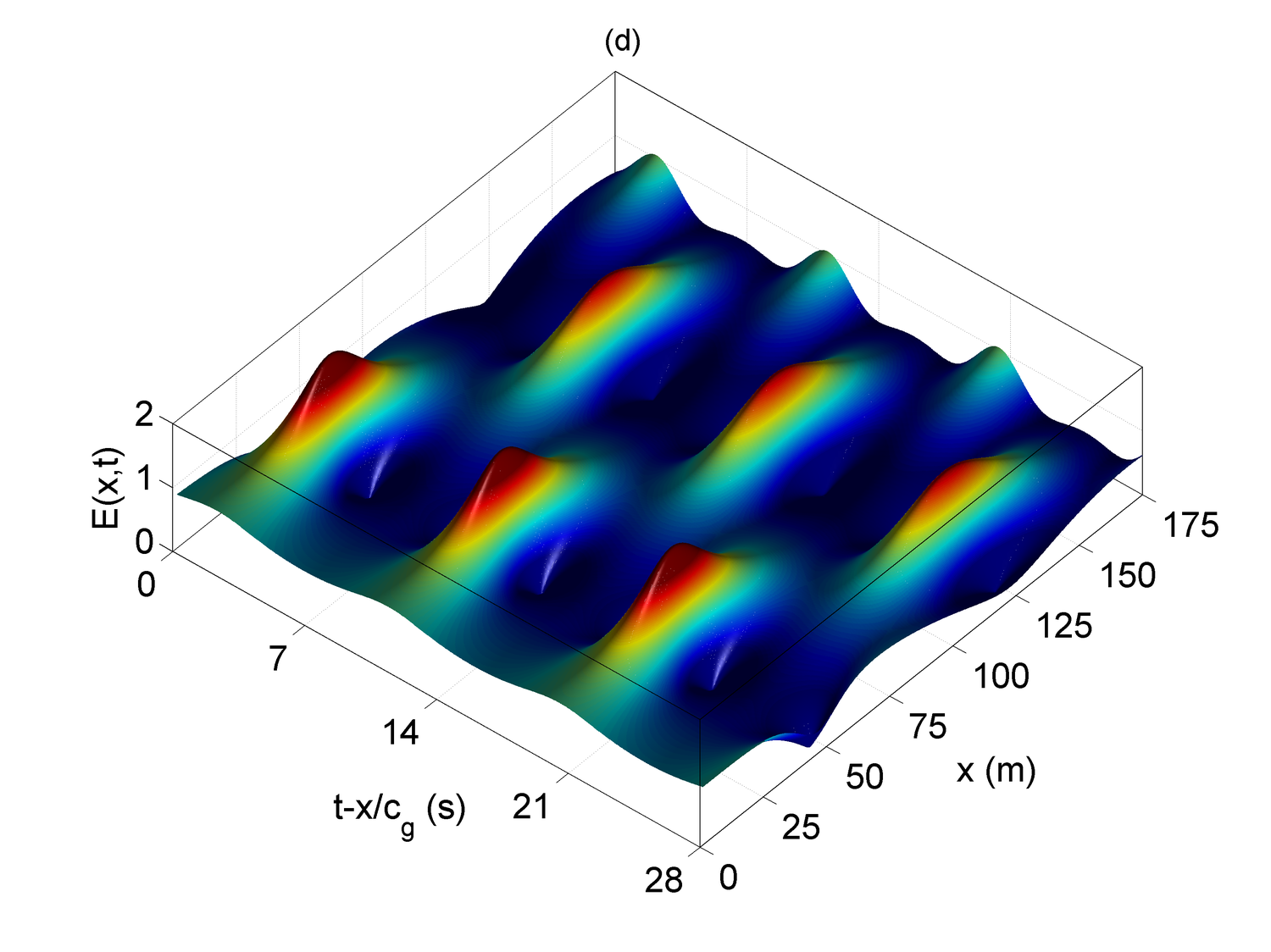}\\
\end{tabular}
\caption{Measured AB envelope along the large wave facility for: (a) the carrier parameters  $a=0.013$ m and $\varepsilon=0.11$, a breather parameter $\mathfrak{a}=0.3$ and dissipation rate $\mathfrak{D}=4.9\,\mbox{x}\,10^{-2}$, (c) the carrier parameters $a=0.023$ m and $\varepsilon=0.13$, a breather parameter $\mathfrak{a}=0.35$ and dissipation rate $\mathfrak{D}=3.9\,\mbox{x}\,10^{-2}$. (b,d) corresponding numerical NLS simulations using the split-step method for the same dissipation rate, breather and carrier parameters as in (a,c) respectively.} 
\label{fig8}
\end{figure}

Before starting experiments to study the influence of the dissipation on hydrodynamic ABs, we determine the corresponding dissipation rate, obviously always naturally existing, when performing experiments in a narrow and one-dimensional water wave basin, in particular, in a significantly long facility. For this purpose, we first generate a regular wave train and measured the attenuation of the wave amplitude during its propagation along the flume. For a regular wave field of amplitude $a=0.013$ m and steepness $\varepsilon=ak=0.11$, where $k$ denotes the wave number, the linear dissipation rate was found to be $\mathfrak{D}=4.9\,\mbox{x}\,10^{-2}$. This corresponds to a wave amplitude attenuation of $25\%$ over a propagation distance in the flume of 155 m. We now use the same parameters for the carrier in order to excite AB on the corresponding background. We set the breather parameter to be $\mathfrak{a}=0.3$. The collected temporal surface elevations are then aligned by the value of the group velocity $c_g$ for comparison with numerical simulations. 
Due to the steepness of the background wave train, a nonlinear correction of the group velocity is needed. The nonlinear group velocity $c_g=\partial\omega/\partial k$ is deduced from the nonlinear relation dispersion given up to the third order:   
$$\omega^2=gk\left(1+\frac 12 \varepsilon^2\right)$$
The envelope of the measured wave trains are then extracted by use of the Hilbert transform. Fig. \ref{fig8}(a) shows the results. 
Clearly, the breather recurrent cycles can be observed in agreement with numerical simulations. This proves the orbit jump {\color{blue}\cite{Haller1,Haller2}}, as shown in Fig. \ref{fig5}(c). In order to confirm the influence of dissipation in engendering recurrent shifted localized envelopes, we performed another set of experiments, however, for different carrier and AB parameters, respectively. In the following we set the amplitude to be $a=0.023$ m, the steepness to be $\varepsilon=0.13$ and the AB parameter $\mathfrak{a}=0.35$, while the dissipation rate is in this case $\mathfrak{D}=3.9\,\mbox{x}\,10^{-2}$. All these chosen parameters are beyond breaking thresholds of the unstable waves, and the parameter $\mathfrak{a}$ is below the threshold of higher order modulation instability {\color{blue}\cite{Erkintalo2}}. The evolution of the wave envelope is shown in Fig. \ref{fig8}(c). The corresponding numerical damped NLSE simulations of both latter observations are shown in Fig. \ref{fig8}(b) and Fig. \ref{fig8}(d) respectively.  
Again, we can here clearly observe the $\pi/2$-phase shift in each cycle of envelope compression. In both cases, numerical simulations are indeed in strong agreement. 
\section{SUMMARY}
We have shown that weak dissipation may engender specific phase-shifted cycles of $\pi/2$ in the evolution of periodic modulationally unstable waves, described by AB. In fact, we discussed the effect of dissipation on the orbit jumps in the complex phase space. The numerical results starting from exact AB initial conditions have been validated by laboratory experiments, conducted in super tank for different set of breather and background parameters. We predict that this novel phenomena, we refer to as phase-shifted FPU recurrence, will motivate a wide range of applications, due to the multidisciplinary nature of the problem, since the NLSE, accurately models the dynamics of hydrodynamic, electromagnetic as well as plasma waves. 
It is worth mentioning that the observation of many breathing cycles is a very challenging task, in particular to study the effect of perturbations on FPU recurrence without the impact of propagation losses. Indeed, overcoming experimental restrictions in hydrodynamics, such as the dissipation or the nonlinear length to name a few, is not something obvious, even for other physical systems. For instance, breather waves have also been intensively studied in optical fibers, but limited to 1 or 2 cycles of recurrence due to the typical normalized dissipation rate $\mathfrak{D}$ estimated about $3.8\,\mbox{x}\,10^{-2}$ (for typical parameters SMF-28 optical fiber and 1-W continuous wave power {\color{blue}\cite{Kibler2,Erkintalo2,DudleyKibler}}). This clearly points out both the extreme importance of the hydrodynamic results reported here and capabilities of the super wave tank compared to nonlinear optics. To go beyond the frontier in terms of testing nonlinear wave theory, it clearly requires to overcome the current limitations of real physical systems.
\section*{ACKNOWLEDGMENTS} 
O.~Kimmoun acknowledges support from the french-Taiwanese ORCHID program of the Hubert Curien Partnership (PHC). B.~Kibler and A.~Chabchoub acknowledge support from the Burgundy Region (PARI Photcom). A.~Chabchoub acknowledges support from The Association of German Engineers (VDI). A.~Chabchoub is an International Research Fellow of the Japan Society for the Promotion of Science (JSPS). M.~Onorato was supported by MIUR grant no. PRIN 2012BFNWZ2. M.Onorato thanks Dr B.~Giulinico for fruitful discussions. 
\\
\appendix
\newenvironment{myequation}{%
\addtocounter{equation}{-1}
\refstepcounter{defcounter}
\renewcommand\theequation{A\thedefcounter}
\begin{equation}}
{\end{equation}}
\section*{APPENDIX : METHODS}
\subsection{Theoretical preliminaries.} 
The evolution of dimensional deep-water packets $\Psi\left(x,t\right)$, propagating in space with the group velocity can be modeled by the deep-water NLSE {\color{blue}\cite{Zakharov}}
\begin{myequation}
\textnormal{i}\left(\Psi_x+\dfrac{1}{c_g}\Psi_t\right)-\dfrac{1}{g}\Psi_{tt}-k^3\left|\Psi\right|^2\Psi=\textnormal{i}\gamma\,\Psi\tag{A}
\label{NLSEdim}
\end{myequation}
Here, $g$ is the gravitational acceleration and $\gamma$ the dissipation rate. The wave number and the wave frequency are connected through the linear dispersion relation $\omega^2=gk$. The group velocity is then equal to $c_g=\dfrac{\omega}{2k}$ \\
\subsection{Towards experimental initials conditions.} 
The experiments have been conducted in deep-water conditions. Considering the depth being $h=1.35$, the parameters of the carrier wave have been chosen accordingly. Once the amplitude $a$ and the steepness $\varepsilon=ak$ being fixed, the wave frequency can be derived from the dispersion relation for deep-water. In the next step, the AB solution (\ref{AB}) from the scaled NLSE (1) have to be transformed accordingly by setting $\xi=-\dfrac{1}{2}a^2k^3 x$, $\tau=\dfrac{\sqrt{2}ak\omega}{2}\left(t-\dfrac{x}{c_g}\right)$ and $\psi=\dfrac{\Psi}{a}$. 
Then, the non dimensional form of the NLSE is given by:
\begin{myequation}
\textnormal{i}\psi_{\xi}+\psi_{\tau\tau}+2|\psi|^2\psi=-\textnormal{i}\dfrac{2\gamma}{a^2k^3}\psi=-\textnormal{i}\mathfrak{D}\psi\tag{B}
\label{NLSEadim}
\end{myequation}
After fixing the breather parameter $\mathfrak{a}$ the Akhmediev-type surface elevation is then given by
\begin{myequation}
\begin{array}{ll}
\eta_{AB}\left(x,t\right)& = \textnormal{Re}\left(\Psi\left(x,t\right)\exp\left[\textnormal{i}\left(kx-\omega t\right)\right]\right.\\
& +\dfrac{1}{2}k\Psi^2\left(x,t\right)\exp\left[2\textnormal{i}\left(kx-\omega t\right)\right]) \tag{C}
\label{ABsurface}
\end{array}
\end{myequation}
The boundary condition of an Akhmediev breather in order to begin the dynamics of this solution is determined by evaluating Eq. (\ref{ABsurface}) at a specific position $x^*$ of interest. 
\subsection{Reconstruction of the wave envelope dynamics.} 
Wave gauges measurements collect the temporal evolution of the water surface at the . The temporal variation of envelope $\Psi(x^*,t)$ can be easily reconstructed from a surface measurement $\eta\left(x^*,t\right)$, by use of the Hilbert transform {\bf H} as the following
\begin{myequation}
\begin{array}{ll}
\Psi\left(x^*,t\right) & =\eta\left(x^*,t\right)+\textnormal{i}\textnormal{\bf{H}}\left[\eta\left(x^*,t\right)\right]\\
& =\eta\left(x^*,t\right)+\dfrac{\textnormal{i}}{\pi}\int_{-\infty}^{+\infty}{\dfrac{\eta\left(x^*,\zeta\right)}{t-\zeta}\textnormal{d}\zeta}\tag{D}
\end{array}
\end{myequation}
Figs. \ref{fig8}(a-b) show the whole temporal evolution of the wave envelope, while evolving in space. The latter have been obtained by interpolating the several Hilbert transformation envelope measurements accordingly. 


\begin{thebibliography}{99}
\bibitem{FPU} E. Fermi, J. Pasta, and S. Ulam, Studies of the Nonlinear Problems, Los Alamos Report LA-1940, (1955).
\bibitem{OnoratoPNAS} M. Onorato, L. Vozella, D. Proment, and Y. V. Lvov, Route to thermalization in the $\alpha$-Fermi–Pasta–Ulam system, {\color{blue}Proceedings of the National Academy of Sciences {\bf112}, 4208 (2015)}. 
\bibitem{Zabusky} N. J. Zabusky and M. D. Kruskal, Interactions of solitons in a collisionless plasma and the recurrence of initial states. {\color{blue}Phys. Rev. Lett. {\bf15},  240 (1965)}.
\bibitem{Newell} D. J. Benney and A. C. Newell, The Propagation of Nonlinear Wave Envelopes, {\color{blue}J. Math. Phys. (N.Y.) {\bf46}, 133 (1967)}.
\bibitem{Zakharov} V. E. Zakharov, Stability of periodic waves of finite amplitude on a surface of deep fluid. {\color{blue} J. Appl. Mech. Tech. Phys. \textbf{9}, 190 (1968)}.
\bibitem{YuenLake1982}H. C. Yuen and B. M. Lake, Nonlinear dynamics of deep-water gravity waves, {\color{blue} Adv. Appl. Mech. {\bf 22}, 67 (1982)}.
\bibitem{Waseda} M. P. Tulin and T. Waseda, Laboratory observations of wave group evolution, including breaking effects, {\color{blue} J. Fluid Mech. {\bf378}, 197 (1999)}.
\bibitem{VanSimaeys} G. Van Simaeys, P. Emplit, and M. Haelterman, Experimental demonstration of the Fermi-Pasta-Ulam recurrence in a modulationally unstable optical wave, {\color{blue} Phys. Rev. Lett. 87, 033902 (2001)}. 
\bibitem{Kibler2} B. Kibler, J. Fatome, C. Finot, G. Millot, G. Genty, B. Wetzel, N. Akhmediev, F. Dias, and J. M. Dudley, Observation of Kuznetsov-Ma Soliton Dynamics in Optical Fibre, {\color{blue} Sci. Rep. {\bf2}, 463 (2012)}.
\bibitem{Mussot} A. Mussot, A. Kudlinski, M. Droques, P. Szriftgiser, and N. Akhmediev, Fermi-Pasta-Ulam Recurrence in Nonlinear Fiber Optics: The Role of Reversible and Irreversible Losses, {\color{blue} Phys Rev. X {\bf 4},  011054 (2014)}.
\bibitem{Bendahmane2} A. Bendahmane, A. Mussot, P. Szriftgiser, O. Zerkak, G. Genty, J. M. Dudley, and A. Kudlinski, Experimental dynamics of Akhmediev breathers in a dispersion varying optical fiber, {\color{blue} Opt. Lett. 39, 4490 (2014)}.
\bibitem{Erkintalo2} M. Erkintalo, K. Hammani, B. Kibler, C. Finot, N. Akhmediev, J. M. Dudley, and G. Genty, Higher-order modulation instability in nonlinear fiber optics, {\color{blue} Phys. Rev. Lett. 107, 253901 (2011)}.
\bibitem{AkhmedievBook} N. Akhmediev and A. Ankiewicz, {\it Solitons: Nonlinear Pulses and Beams} (Chapman \& Hall, London, 1997).
\bibitem{Akhmediev1} N. Akhmediev, V. M. Eleonskii, and N. Kulagin, Generation of periodic sequence of picosecond pulses in an optical fibre: Exact solutions, {\color{blue} J. Exp. Theor. Phys. {\bf 61}, 894 (1985)}.
\bibitem{Abramowitz} M. Abramowitz and I. A. Stegun, \textit{Handbook of Mathematical Functions} (National Bureau of Standards, Applied Mathematics Series {\bf55}, 1964).
\bibitem{Kuznetsov} E. A. Kuznetsov, Solitons in a Parametrically Unstable Plasma, {\color{blue} Sov. Phys. Dokl. {\bf22}, 507 (1977)}.
\bibitem{Ma} Y. C. Ma, The Perturbed Plane Wave Solutions of the Cubic Nonlinear Schrödinger Equation, {\color{blue} Stud. Appl. Math. {\bf60}, 43 (1979)}.
\bibitem{Peregrine} D. H. Peregrine, Water waves, nonlinear Schr\"odinger equations and their solutions. {\color{blue} J. Aust. Math. Soc. Ser. B \textbf{25}, 16 (1983)}. 
\bibitem{OnoratoReview} M. Onorato, S. Residori, U. Bortolozzo, A. Montina, and F.T. Arecchi, Rogue waves and their generating mechanisms in different physical contexts, {\color{blue} Phys. Rep. {\bf528}, 47 (2013)}.
\bibitem{DudleyReview} J. M. Dudley, F. Dias, M. Erkintalo, and G. Genty, Instabilities, breathers and rogue waves in optics, {\color{blue} Nat. Photonics {\bf8} 755 (2014)}.
\bibitem{Kibler} B. Kibler, J. Fatome, C. Finot, G. Millot, F. Dias, G. Genty, N. Akhmediev, and J. M. Dudley, The Peregrine soliton in nonlinear fibre optics. {\color{blue} Nature Physics \textbf{6}, 790 (2010)}.
\bibitem{Chabchoub1} A. Chabchoub, N. P. Hoffmann, and N. Akhmediev, Rogue wave observation in a water wave tank, {\color{blue} Phys. Rev. Lett. \textbf{106}, 204502 (2011)}.
\bibitem{Bailung} H. Bailung, S. K. Sharma, and Y. Nakamura, Observation of Peregrine Solitons in a Multicomponent Plasma with Negative Ions, {\color{blue} Phys. Rev. Lett. {\bf107}, 255005 (2011)}.
\bibitem{Chabchoub2} A. Chabchoub, B. Kibler, J. M. Dudley and N. Akhmediev, Hydrodynamics of periodic breathers, {\color{blue} Phil. Trans. R. Soc. A {\bf372}, 20140005 (2014)}.
\bibitem{DudleyKibler} J. M. Dudley, G. Genty, F. Dias, B. Kibler, and N. Akhmediev, Modulation instability, Akhmediev Breathers and continuous wave supercontinuum generation, {\color{blue} Optics Express {\bf 17} 21497 (2009)}.
\bibitem{Hammani} K. Hammani, B. Wetzel, B. Kibler, J. Fatome, C. Finot, G. Millot, N. Akhmediev, and J. M. Dudley, Spectral dynamics of modulation instability described using Akhmediev breather theory, {\color{blue} Opt. Lett. 36, 2140 (2011)}.
\bibitem{Agrawal} G. P. Agrawal, {\it Nonlinear Fiber Optics}, (Academic Press, 2013).
\bibitem{Dysthe} K. B. Dysthe, Note on a modification to the nonlinear Schrodinger equation for application to deep water waves, {\color{blue} Proc. R. Soc. Lond. A {\bf369}, 105 (1979)}.
\bibitem{Hasegawa} A. Hasegawa and M. Matsumoto, Optical Solitons in Fibers (3rd ed., Springer-Verlag, Berlin Heidelberg, 2003).
\bibitem{Haller1} G. Haller and S. Wiggins, Multi-pulse jumping orbits and homoclinic trees in a modal truncation of the damped-forced nonlinear Schrodinger equation, {\color{blue} Physica D {\bf85}, 311 (1995)}.
\bibitem{Haller2} G. Haller, Homoclinic jumping in the perturbed nonlinear Schrodinger equation, {\color{blue} Comms. of Pure and Appl. Math. {\bf52}, 1 (1999)}.
\end{thebibliography}
\end{document}